%% file: aa.tex
\documentclass[longauth]{aa} 

\usepackage{graphicx}
\usepackage{txfonts}
\usepackage{color}
\usepackage{xspace}
\usepackage{siunitx}
\usepackage{amsmath}
\usepackage{amssymb}
\usepackage[table]{xcolor}
\usepackage{dashbox}
\usepackage{framed}
\usepackage{lipsum}
\usepackage{placeins}
\usepackage{stfloats}
\usepackage{hhline}
\usepackage{pifont}

\DeclareSIUnit{\mas}{mas}
\DeclareSIUnit{\arcmin}{arcmin}
\DeclareSIUnit{\degree}{deg}
\DeclareSIUnit{\year}{yr}
\DeclareSIUnit{\msun}{M_{\odot}}
\DeclareSIUnit{\mjup}{M_{\mathrm{Jup}}}

\usepackage{hyperref}

\hypersetup{
        colorlinks=true, 
        breaklinks=true,
        linkcolor=blue, 
        citecolor=blue, 
        filecolor=blue, 
        urlcolor=blue,
        unicode=false, 
        pdftoolbar=true, 
        pdfmenubar=true, 
        pdffitwindow=false, 
        pdfstartview={Fit}, 
        pdftitle={VISIONS - The VISTA Star Formation Atlas}, 
        pdfauthor={Stefan Meingast}, 
        pdfsubject={I. Survey overview},
        pdfcreator={Stefan Meingast}, 
        pdfkeywords={},
        pdfnewwindow=true, 
        pdfdisplaydoctitle=true 
}

\makeatletter
\renewcommand*\aa@pageof{, page \thepage{} of \pageref*{LastPage}}
\makeatother

\newcommand{\cmark}{\ding{51}}
\newcommand{\xmark}{\ding{55}}

\begin{document}

\title{VISIONS: The VISTA Star Formation Atlas}
\subtitle{I. Survey overview\thanks{Based on observations collected at the European Southern Observatory under ESO program 198.C-2009.}}

\author{
        Stefan Meingast\inst{\ref{inst:uni_vie}} \and
        João Alves\inst{\ref{inst:uni_vie}} \and
        Hervé Bouy\inst{\ref{inst:uni_bordeaux}} \and
        Monika G. Petr-Gotzens\inst{\ref{inst:ESO}} \and
        Verena Fürnkranz\inst{\ref{inst:mpia}} \and
        Josefa E. Großschedl\inst{\ref{inst:uni_vie}} \and
        David Hernandez\inst{\ref{inst:uni_vie}} \and
        Alena Rottensteiner\inst{\ref{inst:uni_vie}} \and
        Magda Arnaboldi\inst{\ref{inst:ESO}} \and
        Joana Ascenso\inst{\ref{inst:uni_porto}, \ref{inst:uni_lisboa}} \and
        Amelia Bayo\inst{\ref{inst:ESO}, \ref{inst:uni_valp}} \and
        Erik Brändli\inst{\ref{inst:uni_vie}} \and
        Anthony G.A. Brown\inst{\ref{inst:leiden_obs}}\and
        Jan Forbrich\inst{\ref{inst:uni_herts}} \and
        Alyssa Goodman\inst{\ref{inst:cfa}} \and
        Alvaro Hacar\inst{\ref{inst:uni_vie}} \and
        Birgit Hasenberger\inst{\ref{inst:uni_vie}} \and
        Rainer Köhler\inst{\ref{inst:uni_gsu}} \and
        Karolina Kubiak\inst{\ref{inst:uni_lisboa}} \and
        Michael Kuhn\inst{\ref{inst:uni_herts}} \and
        Charles Lada\inst{\ref{inst:cfa}} \and
        Kieran Leschinski\inst{\ref{inst:uni_vie}} \and
        Marco Lombardi\inst{\ref{inst:uni_milano}} \and
        Diego Mardones\inst{\ref{inst:uni_santiago}} \and
        Laura Mascetti\inst{\ref{inst:ESO},\ref{inst:terma}} \and
        Núria Miret-Roig\inst{\ref{inst:uni_vie}} \and
        André Moitinho\inst{\ref{inst:uni_lisboa}} \and
        Koraljka Mu\v{z}i\'c\inst{\ref{inst:uni_porto}, \ref{inst:uni_lisboa}} \and
        Martin Piecka\inst{\ref{inst:uni_vie}}  \and
        Laura Posch\inst{\ref{inst:uni_vie}} \and
        Timo Prusti\inst{\ref{inst:estec}} \and
        Karla Pe\~na Ram\'irez\inst{\ref{inst:uni_antof}} \and
        Ronny Ramlau\inst{\ref{inst:uni_linz}, \ref{inst:ricam}} \and
        Sebastian Ratzenböck\inst{\ref{inst:uni_vie},\ref{inst:uni_vie_datasci}} \and
        Germano Sacco\inst{\ref{inst:inaf_firenze}} \and
        Cameren Swiggum\inst{\ref{inst:uni_vie}} \and
        Paula Stella Teixeira\inst{\ref{inst:uni_standrews}} \and
        Vanessa Urban\inst{\ref{inst:uni_vie}} \and
        Eleonora Zari\inst{\ref{inst:mpia}} \and
        Catherine Zucker\inst{\ref{inst:stsci}}
        }

\institute{
Universität Wien, Institut für Astrophysik, T\"urkenschanzstrasse 17, 1180 Wien, Austria\label{inst:uni_vie}
\\ \email{stefan.meingast@univie.ac.at}\and
Université de Bordeaux, Lab. d’Astrophysique de Bordeaux, CNRS, B18N, Allée Geoffroy Saint-Hillaire, 33615 Pessac, France\label{inst:uni_bordeaux} \and
European Southern Observatory, Karl-Schwarzschild-Strasse 2, 85748 Garching bei München, Germany\label{inst:ESO} \and
Max-Planck-Institut für Astronomie, Königstuhl 17, D-69117, Heidelberg, Germany\label{inst:mpia} \and
Universidade do Porto, Faculdade de Engenharia, Rua Dr. Roberto Frias, 4200-465 Porto, Portugal\label{inst:uni_porto} \and
Universidade de Lisboa, Faculdade de Ci\^{e}ncias, CENTRA, Ed. C8, Campo Grande, P-1749-016 Lisboa, Portugal\label{inst:uni_lisboa} \and
Universidad de Valpara\'{\i}so, Instituto de F\'{\i}sica y Astronom\'{\i}a, Gran Breta\~na 1111, Valpara\'{\i}so, Chile\label{inst:uni_valp} \and
Leiden University, Leiden Observatory, Niels Bohrweg 2, 2333 CA Leiden, The Netherlands\label{inst:leiden_obs} \and
University of Hertfordshire, Centre for Astrophysics Research, College Lane, Hatfield AL10 9AB, UK\label{inst:uni_herts} \and
Harvard-Smithsonian Center for Astrophysics, Cambridge, MA 02138, USA\label{inst:cfa} \and
The CHARA Array of Georgia State University, Mount Wilson Observatory, Mount Wilson, CA 91023, USA\label{inst:uni_gsu} \and
Università degli Studi di Milano, Dipartimento di Fisica via Celoria 16, I-20133 Milano, Italy\label{inst:uni_milano} \and
Universidad de Chile, Departamento de Astronom\'{\i}a, Las Condes, Santiago, Chile\label{inst:uni_santiago} \and
Terma GmbH, Bratustraße 7, 64293 Darmstadt, Germany\label{inst:terma} \and
European Space Agency (ESA), European Space Research and Technology Centre (ESTEC), Keplerlaan 1, 2201 AZ Noordwijk, The Netherlands\label{inst:estec} \and
Universidad de Antofagasta, Centro de Astronom\'{\i}a (CITEVA), Av. Angamos 601, Antofagasta, Chile\label{inst:uni_antof} \and
Johannes Kepler University Linz, Industrial Mathematics Institute, Altenbergerstraße 69, A-4040 Linz, Austria\label{inst:uni_linz} \and
Johann Radon Institute for Computational and Applied Mathematics (RICAM), Austrian Academy of Sciences (AAS), Altenbergerstr. 69, A-4040 Linz, Austria\label{inst:ricam} \and
University of Vienna, Research Network Data Science at Uni Vienna, Austria\label{inst:uni_vie_datasci} \and
INAF – Osservatorio Astrofisico di Arcetri, Largo E. Fermi, 5, 50125, Firenze, Italy\label{inst:inaf_firenze} \and
University of St Andrews, SUPA, School of Physics \& Astronomy, North Haugh, St Andrews KY16 9SS, UK\label{inst:uni_standrews} \and
Space Telescope Science Institute, 3700 San Martin Dr, Baltimore, MD 21218, USA\label{inst:stsci}
}

\date{Received 22 December 2022 / Accepted 19 January 2023}

\abstract{VISIONS is an ESO public survey of five nearby ($d < 500\,\si{pc}$) star-forming molecular cloud complexes that are canonically associated with the constellations of Chamaeleon, Corona Australis, Lupus, Ophiuchus, and Orion. The survey was carried out with the Visible and Infrared Survey Telescope for Astronomy (VISTA), using the VISTA Infrared Camera (VIRCAM), and collected data in the near-infrared passbands $J$ (\SI{1.25}{\micro\metre}), $H$ (\SI{1.65}{\micro\metre}), and $K_S$ (\SI{2.15}{\micro\metre}). With a total on-sky exposure time of \SI{49.4}{\hour} VISIONS covers an area of \SI{650}{\degree\squared}, it is designed to build an infrared legacy archive with a structure and content similar to the Two Micron All Sky Survey (2MASS) for the screened star-forming regions. Taking place between April~2017 and March~2022, the observations yielded approximately 1.15 million images, which comprise \SI{19}{TB} of raw data. The observations undertaken within the survey are grouped into three different subsurveys. First, the wide subsurvey comprises shallow, large-scale observations and it has revisited the star-forming complexes six times over the course of its execution. Second, the deep subsurvey of dedicated high-sensitivity observations has collected data on areas with the largest amounts of dust extinction. Third, the control subsurvey includes observations of areas of low-to-negligible dust extinction. Using this strategy, the VISIONS observation program offers multi-epoch position measurements, with the ability to access deeply embedded objects, and it provides a baseline for statistical comparisons and sample completeness -- all at the same time. In particular, VISIONS is designed to measure the proper motions of point sources, with a precision of \SI{1}{\mas\per\year} or better, when complemented with data from the VISTA Hemisphere Survey (VHS). In this way, VISIONS can provide proper motions of complete ensembles of embedded and low-mass objects, including sources inaccessible to the optical ESA \textit{Gaia} mission. VISIONS will enable the community to address a variety of research topics from a more informed perspective, including the 3D distribution and motion of embedded stars and the nearby interstellar medium, the identification and characterization of young stellar objects, the formation and evolution of embedded stellar clusters and their initial mass function, as well as the characteristics of interstellar dust and the reddening law.}


\maketitle

\section{Introduction}
\label{sec:introduction}

We are currently witnessing a new golden age of astronomy by virtue of the availability of an impressive array of powerful telescopes stationed both in space and on Earth. The swift technological developments of the last few decades have led to the deployment of increasingly complex astronomical instrumentation. Equipped with such advanced instruments, said telescopes now enable astronomers and data scientists alike to make new discoveries on an unprecedented scale. In consequence, a  topic that has seen tremendous advances over the past decades is related to the formation of stars, a process that regulates cosmic evolution from small to large scales and governs cosmology, galaxy evolution, planet formation, and the origin of life.

Within the field of star formation, we have found certain empirical rules that generally connect the attributes of molecular clouds with the properties of their transformation process to new stars. Yet, despite all advances, we still lack a quantitative theory that, given a limited set of initial conditions and measured properties of molecular clouds, is capable of predicting the star formation rate and the basic physical properties of stars. Arguably, the most critical unsolved problems confronting star formation research today are: understanding star formation in the context of the Galaxy and its components, the origin of the stellar mass distribution, the physical processes that control the star formation rate in molecular gas, and young stellar object (YSO) evolution, together with its implications for planet formation, as well as the origin and evolution of stellar clusters.

The innovations of the last decades have brought a wealth of new data to the community at infrared and (sub-)millimeter wavelengths, where the star formation process can be observed directly. Some of the main contributors to the ever-increasing observational archives are the Very Large Telescope and the Atacama Large Millimeter/submillimeter Array operated by the European Southern Observatory (ESO), as well as the \textit{Gaia} \citep[][]{Gaia}, \textit{Hubble}, \textit{JWST} \citep[][]{jwst}, \textit{WISE} \citep[][]{WISE}, \textit{Spitzer} \citep[][]{Spitzer}, \textit{Herschel} \citep[][]{Herschel}, and \textit{Planck} \citep[][]{PlanckI}, space observatories. With this tremendous amount of available data, an obvious miss in the landscape of currently available astronomical surveys is a deep, wide-field, near-infrared (NIR) high-image quality survey that not only enables the characterization of YSOs and their surrounding environment on a scale of \SI{100}{AU} but, at the same time, it also allows for a global and consistent probing of YSO dynamics. 

With thousands of YSOs distributed over several giant molecular cloud complexes, the solar neighborhood represents one of the regions in our local galaxy that are particularly rich in star formation sites. In a recent study, \citet{Zucker22} indicated that this plethora of nearby star formation sites could  itself be the result of feedback processes originating from a previous generation of young stars. Our proximity to these sites of star formation constitutes a fortunate circumstance, as it is only in these environments that we possess the capability to spatially resolve individual young stars, monitor the properties of gaseous filaments in high resolution, and directly observe feedback mechanisms that regulate the rate at which new stars are born.

\begin{figure*}[p]
 \centering
 \resizebox{\hsize}{!}{\includegraphics[]{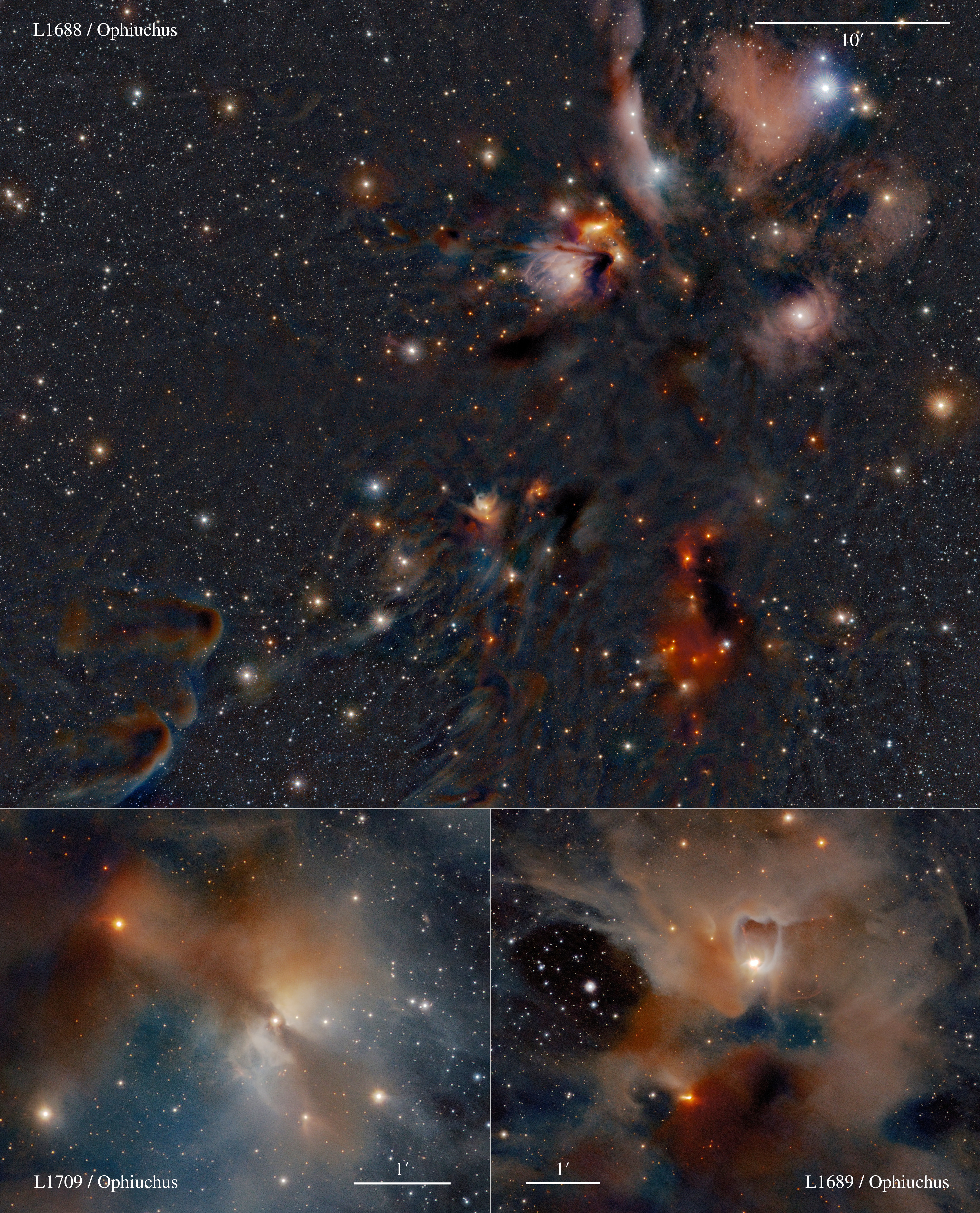}}
 \caption{Ongoing star formation processes as observed by VIRCAM/VISTA in the VISIONS program, using color images assembled from the near-infrared passbands $J$, $H$, and $K_S$. The top panel shows L1688 in the Ophiuchus star-forming complex, depicting several dozen prominent young stellar objects located within one of the closest embedded clusters to Earth. The young stars are surrounded by a highly structured cloud composed of gas and dust that is largely being shaped by young, optically revealed hot stars in the vicinity. The bottom panels further highlight the complex morphology of the gaseous, dusty neighborhood of partly concealed young stars in the Ophiuchus molecular clouds L1709 and L1689.}
 \label{img:matrix1}
\end{figure*}

\begin{figure*}[p]
 \centering
 \resizebox{\hsize}{!}{\includegraphics[]{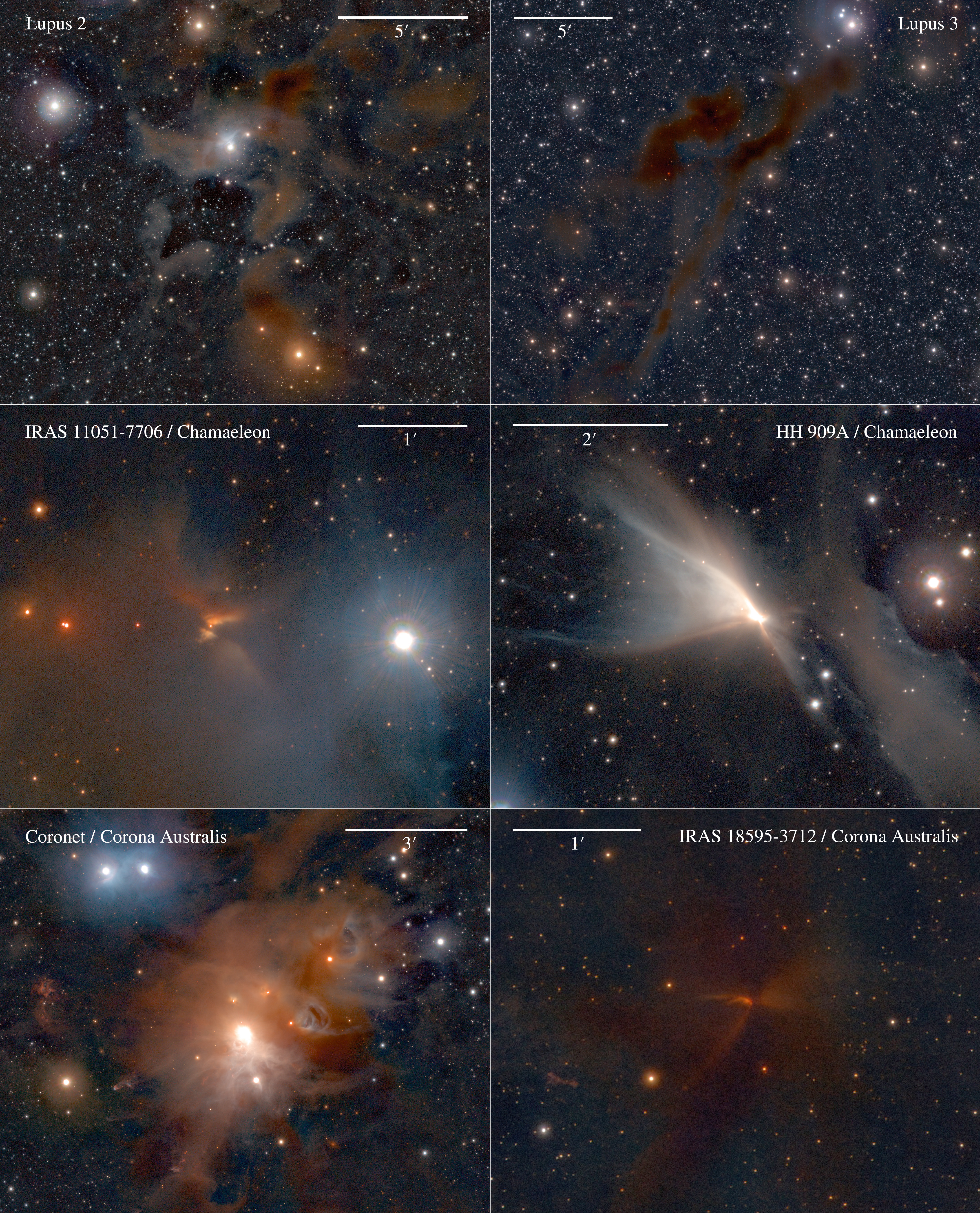}}
 \caption{In the top panels, images of the Lupus 2 and Lupus~3 complexes illustrate the intricate and diverse morphology of dust clouds. The panels in the center portray parts of the Chamaeleon region, which includes young stars that form spectacular cloud shapes, featuring large, cone-shaped reflection cavities. In the bottom-left panel, the Coronet cluster in the Corona Australis star-forming complex is seen illuminating large parts of the surrounding gas and dust. In the bottom right panel, the star-disk-envelope system of a deeply embedded young star becomes visible. By tracing the dust cavity toward the lower left of the image, a faint part of an outflow, ejected from the protostar, can be distinguished.}
 \label{img:matrix2}
\end{figure*}

The VISIONS survey program was established as a natural continuation of the Vienna Survey in Orion project \citep[VISION;][]{Meingast16}, an observing program focused on the characterization of dust properties and the YSO population of the Orion~A molecular cloud. With VISIONS, we provide a survey of five nearby ($d < 500\,\si{pc}$) star-forming complexes that are observable from the southern hemisphere. The names of the regions are derived from the constellations with which they are associated: Chamaeleon, Corona Australis, Lupus, Ophiuchus, and Orion (for an overview of the star-forming regions, see, e.g., \citealp{HandbookI,HandbookII}). In Figs.~\ref{img:matrix1} and \ref{img:matrix2}, we showcase color images of particularly striking objects observed in the VISIONS program, depicting young stars and their dusty surroundings in various shapes and evolutionary stages. VISIONS is aimed at building a legacy NIR archive, thereby extending the data products provided by the Two Micron All Sky Survey \citep[2MASS;][]{2mass_cat,2mass} to probe the contents of the star-forming molecular clouds across the entire stellar and substellar mass spectrum. As such, the survey will also constitute a valuable resources for a new generation of instruments (e.g., MOONS; \citealp{Moons}). Moreover, VISIONS is designed specifically to be complementary to the \textit{Gaia} mission. The survey will connect the \textit{Gaia} astrometric frame to the embedded nearby stellar populations, making it possible to derive proper motions of deeply embedded YSOs to study their dispersal into the galactic field. Being complementary to existing and future observations across the entire electromagnetic spectrum, VISIONS will bring new insights into the processes that govern the transformation of clouds composed of gas and dust into stars and planets.

In this manuscript, we first give a description of the instrumentation and technical details of the survey in Sect.~\ref{sec:description}, before elaborating on the scientific goals in Sect.~\ref{sec:scientific_objectives}. Furthermore, we adopt the terminology established by the European Southern Observatory (ESO). For example, a single snapshot of all detectors of the camera used is called a "pawprint". From a series of such pawprints a contiguous image, called a "tile", can be constructed. Likewise, the integration time for a single snapshot of the camera is called the detector integration time (DIT), which can be stacked NDIT times for one readout.

\section{Survey description}
\label{sec:description}

In this section, we provide a brief overview of the instrumentation used to collect survey data, followed by a detailed description of the design philosophy of the survey, together with its implementation, requirements, and scheduling.

\subsection{Instrumentation}
\label{ssec:instrumentation}

The VISIONS survey uses the Visible and Infrared Survey Telescope for Astronomy \citep[VISTA,][]{vista}, which is integrated into the Cerro Paranal facilities operated by ESO in Chile. Although the VISTA telescope is located on a separate peak from the nearby Very Large Telescope, the typical observing conditions are very similar. The site is ideal for observations at infrared wavelengths and offers photometric observing conditions (no visible clouds; transparency variations less than 2\%, \citealp{Kerber16}) on about one out of four nights, with a median seeing of \SI{0.66}{\arcsec} and precipitable water vapor of \SI{2}{\milli\metre} or lower for \SI{50}{\percent} of the nights\footnote{Information on the astroclimate of the observing site was taken from \href{https://www.eso.org/sci/facilities/paranal/astroclimate/site.html}{www.eso.org/sci/facilities/paranal/astroclimate/site.html}}.

With a \SI{4}{\meter} aperture, VISTA delivers an f/3.25 beam at the Cassegrain focus. There, the VISTA infrared camera \citep[VIRCAM;][]{vircam} images a field of view with \SI{1.65}{\degree} in diameter. The camera houses sixteen 2k~$\times$~2k Raytheon VIRGO HgCdTe detectors with a typical quantum efficiency of 90 -- \SI{95}{\percent} between 1 and \SI{2.5}{\micro\metre}. The detectors are arranged in a sparse 4~$\times$~4 pattern and individually cover a $11.6~\times~11.6\,\si{arcmin}$ field with a mean pixel scale of \SI{0.339}{\arcsec \per pixel}. The intradetector gaps measure \SI{~11.4}{\arcmin} and \SI{~4.9}{\arcmin} in the X and Y axis of the system, respectively. As a result, the camera captures an area of about \SI{0.6}{\degree\squared}, albeit with large gaps between individual images, with a single exposure. Consequently, the construction of a fully contiguous tile requires a set of at least six individually shifted pawprints. Across a co-added tile, this results in differences in field coverage, total exposure time, and exact observation date. The effective exposure of a tile can be calculated with $t_{\mathrm{exp}} = 2 \times \mathrm{DIT} \times \mathrm{NDIT} \times \mathrm{NJITTER}$, where NJITTER is the number of jittered positions per offset. The factor of 2 in this equation is a consequence of the offset pattern in which, disregarding additional jitter positions, most of the areas of a tile are observed twice. The process of jittering is used in the interest of enhancing the quality of the post-processed data: The detectors show a large number of bad pixels (up to \SI{2}{\percent}), many of which appear clustered rather than randomly distributed. To facilitate the effective rejection of bad pixels during co-addition, a jitter pattern can be executed at each offset position. It should also be noted that the detectors suffer from non-linearity of the order of up to a few percent at \SI{10000}{ADU}. 

VISIONS is one of the seven ESO VISTA cycle 2 public surveys \citep[][]{Arnaboldi19}. The observations are executed in service mode (also referred to as queue scheduling mode) to collect data in the (NIR) $J, H, \mathrm{and}$ $K_S$ passbands, with mean wavelengths of 1.25, 1.65, and \SI{2.15}{\micro\meter} and a full width at half maximum (FWHM) of 0.17, 0.29, and \SI{0.31}{\micro\meter}, respectively. For more detailed information on the telescope and camera, we refer the reader to the numerous resources available on the ESO website and specifically to the VIRCAM user manual\footnote{\href{https://www.eso.org/sci/facilities/paranal/instruments/vircam.html}{www.eso.org/sci/facilities/paranal/instruments/vircam.html}}.

\subsection{Survey setup}
\label{ssec:survey_setup}

\begin{figure*}
        \centering
    \resizebox{\hsize}{!}{\includegraphics[]{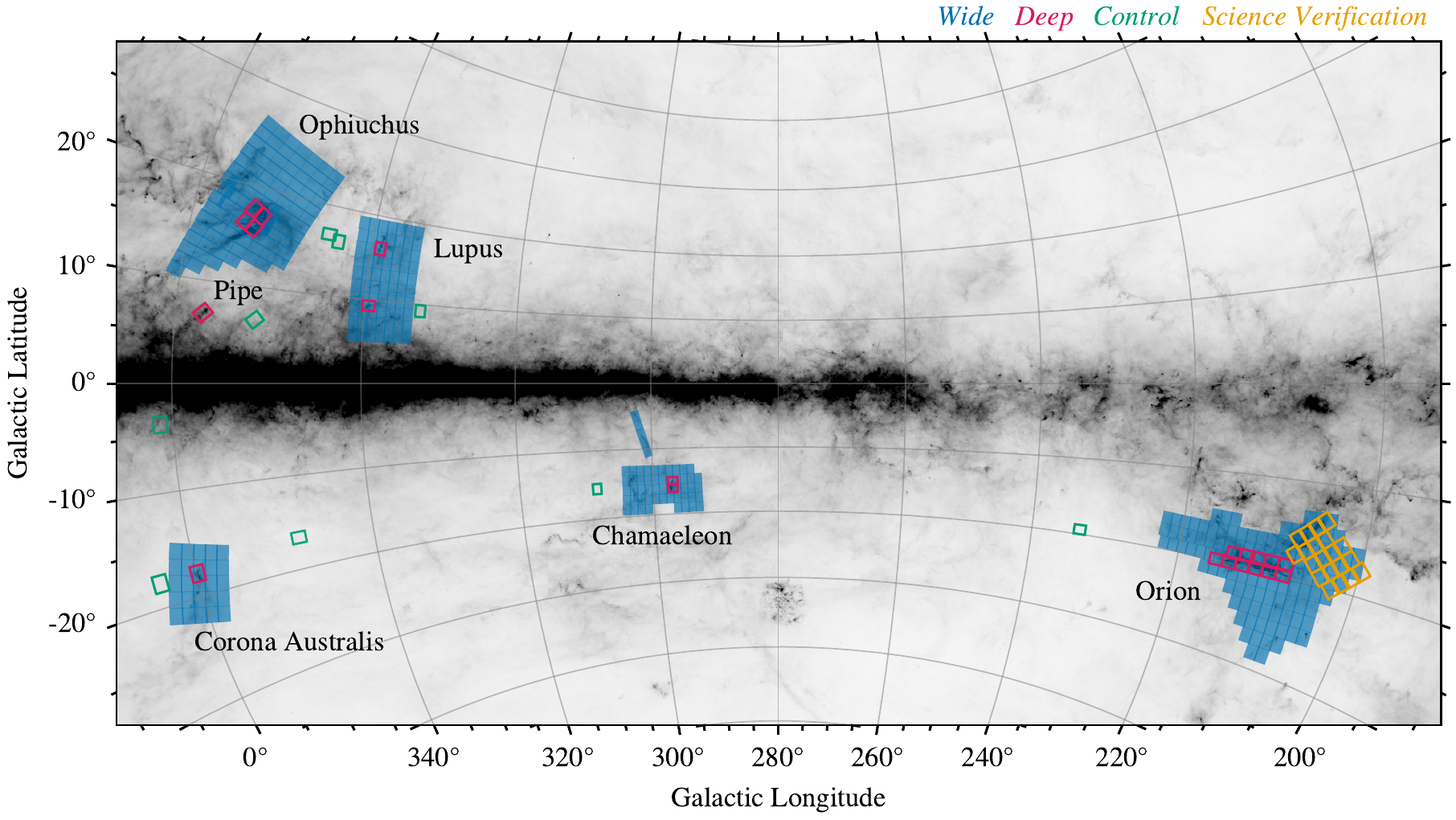}}
        \caption{Star-forming regions observed in the VISIONS survey plotted on top of \textit{Planck} data at \SI{857}{\giga\hertz}. The individual colors of the marked regions portray the subsurveys of the program. The filled blue tiles represent the coverage of the wide program, whereas the red and green rectangles indicate the on-sky locations of data collected as part of the deep and control observations, respectively. Science verification data, collected in the region of the Orion star forming complex, is depicted in orange.}
        \label{img:coverage}
\end{figure*}

To meet the scientific goals of the VISIONS survey (Sect.~\ref{sec:scientific_objectives}) several aspects must be considered, as not all science cases can be addressed with a single observation strategy. For instance, studies on the embedded stellar populations have different observation requirements than a survey with the aim of measuring proper motions. Hence, for regions without a significant amount of extinction, shallow and relatively fast observation sequences can deliver high-quality, multi-epoch data for large areas on the sky. On the other hand, parts of the clouds exhibiting large column densities require longer integration times to both reach deeply embedded sources and observe a sufficiently large number of background stars to construct well-sampled reddening maps at the same time. To achieve an efficient distribution of the available observing time, the VISIONS program is divided into three independent but complementary subsurveys, namely: the wide, deep, and control subsurveys. A summary of the locations, the observation setup, and other metrics for each region can be found in Table~\ref{tab:overview}.

\input{table_setup}

The targeted star-forming regions span sizes on the order of tens of parsecs. However, since they are relatively near to Earth (100 -- $400$~\si{pc}), their contiguous gas distributions cover hundreds of square degrees in the sky. VIRCAM images an area of about \SI{1.8}{\degree\squared} on a minimum of two different pixels with an observing sequence that includes the six offsets and at least one additional jitter position per offset. Yet, even with such a large field of view, several hundred tiles are required to cover the areas of interest. Figure~\ref{img:coverage} shows the on-sky coverage of all subsurveys, superimposed on a view of the Milky Way as seen by \textit{Planck} at \SI{857}{\giga\hertz}\footnote{Data obtained from the \textit{Planck} Legacy Archive, which is accessible at \href{http://pla.esac.esa.int/pla}{pla.esac.esa.int/pla}}, in a stereographic projection centered on $l=280\,\si{\degree}$, and $b=0\,\si{\degree}$. For each region, the figure depicts wide, deep, and control fields in blue, red, and green, respectively. As described below, the deep and control fields for Orion are not part of VISIONS itself, as these data have already been collected, processed, and published by \citet{Meingast16}. The yellow boxes mark the VISTA science verification fields, which cover large parts of the Orion~B molecular cloud complex \citep[e.g.,][program ID: 60.A-9285]{Petr-Gotzens11}. 

\subsubsection{Wide subsurvey}
\label{sssec:wide_subsurvey}

\begin{figure*}
        \centering
    \resizebox{\hsize}{!}{\includegraphics[]{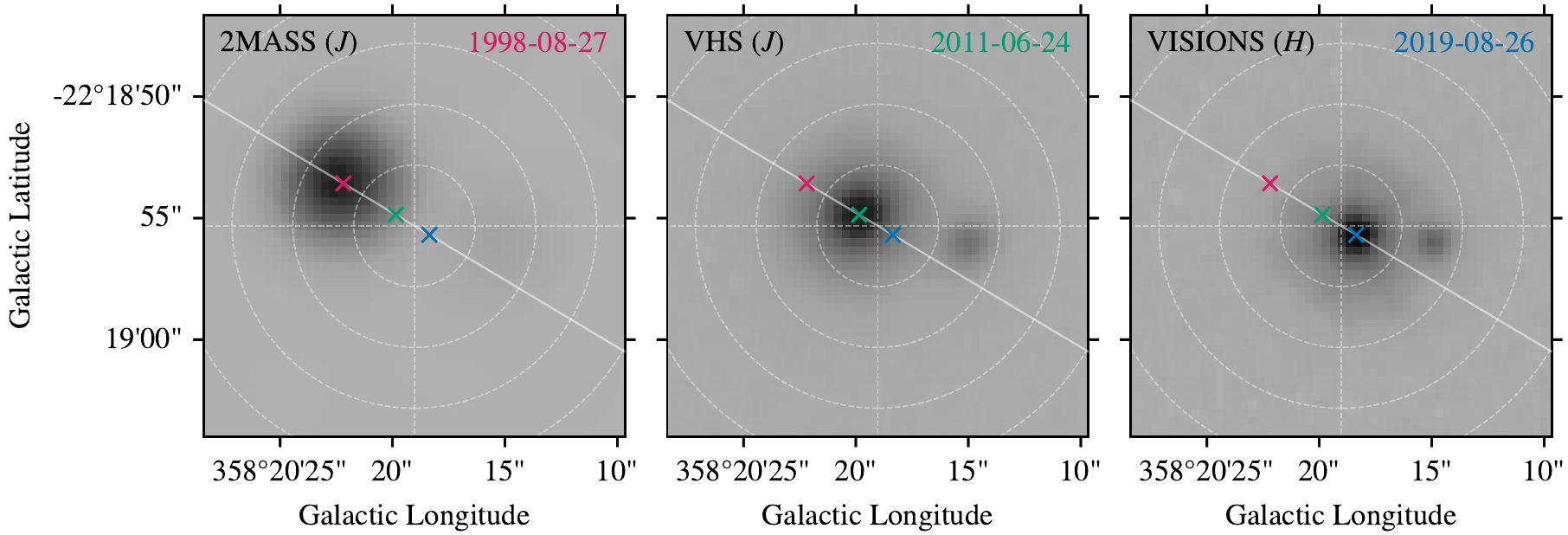}}
        \caption{Multi-epoch imaging data for the high proper motion star 2MASS 19205383-3950043. The leftmost panel shows 2MASS observations captured in 1998, whereas the middle panel displays data from the VHS survey in the $J$ passband from 2011 and the rightmost panel shows VISIONS $H$-band data collected in August 2019. The crosses mark the star's position in each epoch and the white solid line depicts its trajectory. Also visible in the VHS and VISIONS images is another source with an offset toward the Galactic West of about \SI{5}{\arcsec}, which is not discernible in 2MASS data due to the shallower sensitivity limit of the survey.}
        \label{img:proper_motion}
\end{figure*}

The goal of the wide survey is to provide data for large areas over the course of six epochs, which will facilitate proper motion measurements for sources beyond the sensitivity limit of the \textit{Gaia} mission. We selected different survey areas for each region to cover their distribution of gas and young stars. For Ophiuchus, the survey covers \SI{159}{\deg\squared}, comprising the greater Rho Ophiuchi cloud complex with the molecular clouds Barnard~42 through 47, 51, 57, 229, and 238. Furthermore, the field extends several degrees to the Galactic North, also including the Upper Scorpius OB association \citep{dezeeuw99}. The Lupus wide field measures \SI{115.9}{\deg\squared} and contains the molecular clouds Lupus~I through IV. In the Galactic South, this field borders the VVVX public survey limit at $b=5\,\si{\deg}$ \citep{vvv, vvvx}. In the Corona Australis region, VISIONS encompasses \SI{41.4}{\deg\squared}, containing the Coronet cluster and the surrounding low column-density gas. In Chamaeleon, a contiguous field of about \SI{84.5}{\deg\squared} connects observations for the clouds Chamaeleon I, II, and III, while a separate field (\SI{~9}{\deg\squared}) images the Musca filament, which is situated close to the Galactic plane. Orion constitutes the largest wide survey area with \SI{188.8}{\deg\squared}. This field comprises the giant molecular clouds Orion~A and Orion~B, as well as further outlying, oftentimes cometary-shaped clouds, found far to the Galactic east and south. In addition to the regular wide fields, which have been observed six times over the $\sim$ $5\,\si{yr}$ duration of the survey program, we have scheduled an additional epoch at the very end of the VISIONS program for the deep subsurvey pointings. This addition, called Epoch X, guarantees the longest possible time baselines for the deeply embedded population. Summing all individual fields, the wide survey area measures \SI{598.4}{\deg\squared}, for which we have requested observations of a total of \SI{2286} tiles, comprising \SI{68580} VIRCAM pawprints or \SI{1097280} individual images.

\begin{figure*}
        \centering
    \resizebox{\hsize}{!}{\includegraphics[]{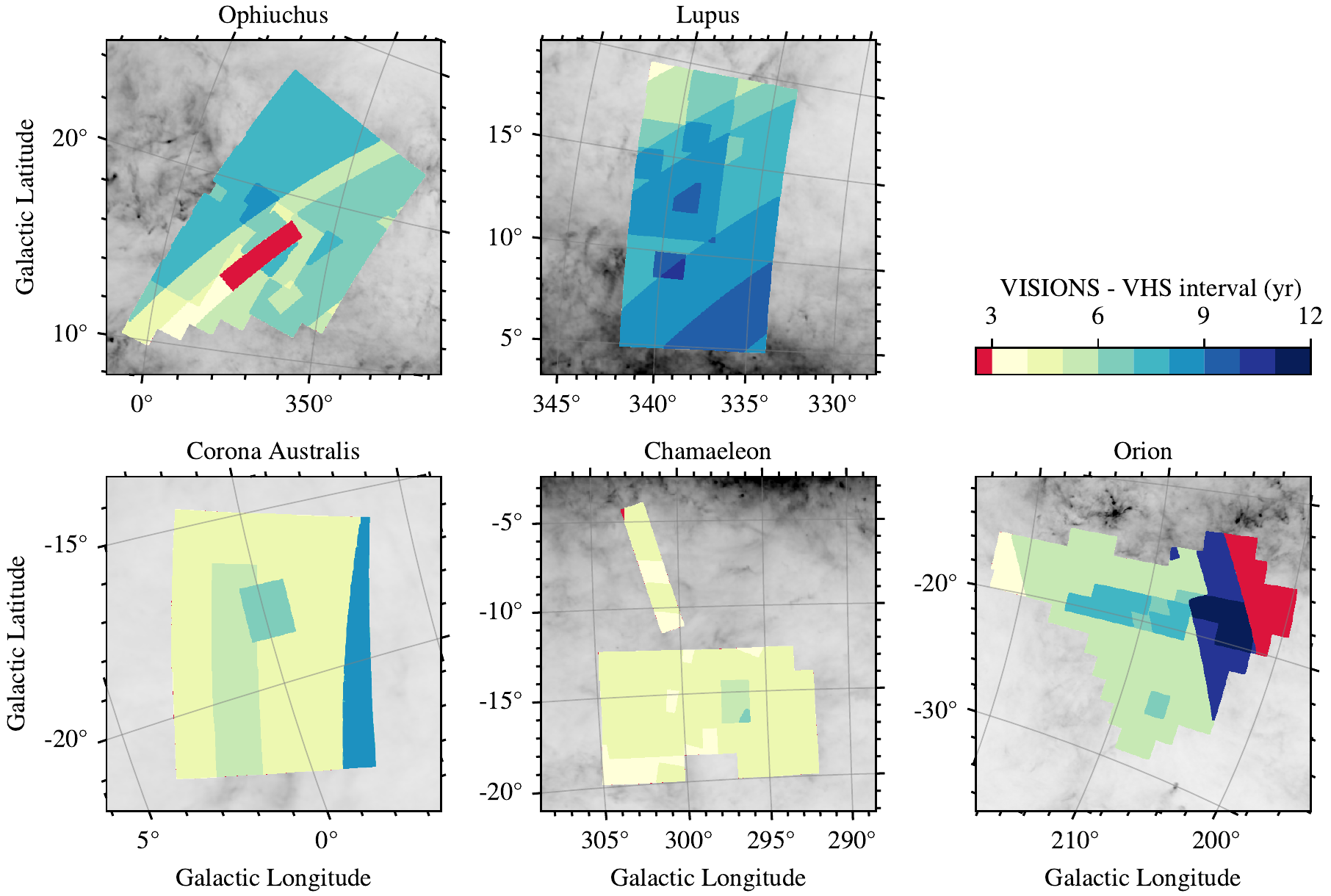}}
        \caption{Maximum available time baseline between VISIONS and VHS observations. Each panel shows a close-up view of the star-forming complexes, as imaged in the wide subsurvey, on top of a \textit{Planck} \SI{857}{\giga\hertz} map. The colorscale shows the time difference between the first VHS observations and the last VISIONS wide field, mapped onto a healpix grid with a pixel size of \SI{47}{arcmin^2}. Time intervals of less than \SI{3}{yr} are indicated in red. A distinct pattern becomes visible, depicting the complex observation strategy of both surveys.}
        \label{img:vhs_baseline_sky}
\end{figure*}

The images for this subsurvey were taken exclusively in the $H$ passband in order to complement the VISTA Hemisphere Survey \citep[VHS;][ESO program ID 179.A-2010]{vhs}, a first-generation VISTA public survey. Over its runtime from 2009 to 2022, VHS mapped the entire southern hemisphere (DEC$~ <0~\si{\deg}$) in the $J$ and $K_S$ passbands, with an effective tile exposure time of \SI{60}{\second}. Combined, the VISIONS wide and VHS programs can provide the full set of NIR passbands, which are already well known from 2MASS. Furthermore, similarly to the VHS data collection strategy, we have favored relatively short integration times to efficiently cover large parts of the star-forming regions over multiple epochs. Consequently, a single tile of our wide survey has a total effective exposure time of \SI{60}{\second} and is built from from five jittered positions at each offset, with DIT$=3~\si{\second}$ and NDIT$=2$. However, in contrast to VHS, which includes only two jittered positions with longer DITs, we have preferred a greater number of individual images over longer single integration times, to ensure a proper rejection of bad pixels during co-addition. Altogether, the wide subsurvey constitutes the largest part of the VISIONS program, with a total of \SI{36.1}{\hour} of on-sky exposure time.

In addition to providing photometry in the $J$ and $K_S$ passbands, VHS also presents a significant extension of the observed time baseline, a parameter highly relevant for measuring proper motions. As a first-generation VISTA public survey, VHS began collecting data in the VISIONS fields during ESO Period 85, with the first observations executed on March 23, 2010 in Lupus. As a consequence, complementing VISIONS with VHS potentially offers a baseline of more than \SI{10}{\year}, facilitating high-precision proper motion measurements. For bright sources, this baseline can even be extended by another \SI{10}{\year} by including observations from 2MASS. To illustrate the effects of combining VISIONS with the two older surveys to create longer baselines, Figure~\ref{img:proper_motion} shows the case of the star 2MASS 19205383-3950043, for which \textit{Gaia} has determined a proper motion of \SI{196.41}{\mas\per\year}. From left to right, the panels in the figure indicate the position of the star as imaged by 2MASS, VHS, and VISIONS, respectively. The star's position in each epoch, provided by \textit{Gaia}, is marked with a cross whose color corresponds to the indicated observing date in the top right corner of the respective panel. The solid white line marks the trajectory computed from \textit{Gaia} astrometry. The location of the source in each image aligns well with the \textit{Gaia} measurements and its movement along the projected trajectory is clearly visible across the different frames.

As data collection for VHS continued well into the phase of the second-generation VISTA public surveys, the interval between VISIONS and VHS observations is significantly shortened for large parts of the observed fields. Figure~\ref{img:vhs_baseline_sky} shows the sky coverage of VISIONS in a healpix grid with a pixel size of \SI{47}{arcmin^2} (resolution order 9), where the color scale depicts the maximum available time interval between VISIONS and VHS. The figure reveals a complex pattern in time baselines caused by the discrete and often irregular observing schedule of both public surveys. The spatial distribution of time intervals shows a characteristic pattern aligned along lines of constant declination as a consequence of the VHS observation sequence. In addition, superimposed on this global pattern, some individual tiles become visible. These are caused, for instance, by wide epoch X on the VISIONS deep fields or necessary repetitions due to poor data quality. The red areas in the Ophiuchus and Orion regions have VHS-VISIONS baselines that span less than \SI{3}{yr}. Regarding these regions, it should be noted that at the time of the VISIONS observations, the corresponding fields in Ophiuchus were not available from VHS, most likely because of difficulties associated with the limitation of available stars for the autoguider and active optics CCDs. As a consequence, we have included additional observation blocks to collect these missing data in the $J$ and $K_S$ passbands, along with our usual $H$-band observations. In Orion, the fields marked in red lie above the declination limit for VHS ($\delta$ < \SI{0}{deg}), however, the corresponding areas are still included in the wide observations, as this area has already been covered with VIRCAM during the phase of VISTA science verification in 2009.

\begin{figure}
        \centering
    \resizebox{\hsize}{!}{\includegraphics[]{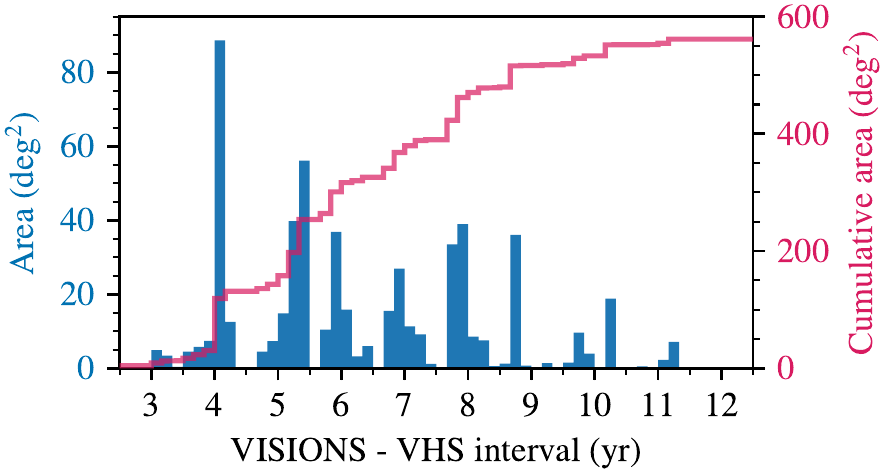}}
        \caption{Distribution of the maximum available time interval between VHS and VISIONS observations, constructed from the healpix grid in Fig.~\ref{img:vhs_baseline_sky}. The blue histogram shows the captured area as a function of the available time difference between the surveys. The red line displays the same statistic in cumulative form. As a consequence of the broad, multi-modal range of available time baselines, the accuracy of derived proper motions will vary accordingly.}
        \label{img:vhs_baseline_hist}
\end{figure}

Figure~\ref{img:vhs_baseline_hist} displays the VISIONS-VHS time baseline of the healpix grid shown in Fig.~\ref{img:vhs_baseline_sky} as a histogram. The blue-filled histogram shows the covered area in \si{\deg\squared} as a function of the time interval between VISIONS and VHS. The red line represents the same data in cumulative form\footnote{The cumulative sum of the area in Fig.~\ref{img:vhs_baseline_hist} does not match the values listed in Table~\ref{tab:overview} because a patch in Orion lies outside the area covered by VHS and is therefore not counted in the histogram.}. The multi-modal distribution in the blue histograms is derived from the visibility of the regions, which is in turn dictated by the discrete timings of the VISIONS wide schedule and the priorities assigned to the individual observation blocks. The histograms reveal that about \SI{150}{\deg\squared} of the observed star-forming regions have a maximum time interval of only \SI{4}{yr} or less. Most of these areas are associated with the Corona Australis and Chamaeleon fields. Furthermore, we find that for half of the survey area the time interval is approximately \SI{6}{yr} or less, with only a relatively small fraction possessing baselines of more than \SI{10}{yr}. As a consequence, we expect the precision of future proper motion measurements to vary significantly throughout the various survey fields.

\subsubsection{Deep subsurvey}
\label{sssec:deep_subsurvey}

The deep subsurvey of the VISIONS program targets areas characterized by high column densities and aims to deliver highly sensitive observations, penetrating deeply into the molecular clouds. For a precise field selection, we inspected the local column density distribution in dust emission measured by Planck at \SI{857}{\giga\hertz}, before arranging the field coverage so that parts with high extinction and ongoing star formation are positioned within the selected observed areas. For Ophiuchus, we included four adjacent tile positions in a $2 \times 2$ arrangement, covering, among others, the prominent dark clouds L1688, L1689, and L1709 associated with the Rho Ophiuchi cloud complex \citep{Lynds62}. In Lupus, two spatially separate tiles focus on the clouds Lupus~I and III, called Lupus North and Lupus South in the VISIONS deep program. For Corona Australis, we observe the Coronet star cluster together with the surrounding dense gas and in Chamaeleon two North-South aligned tiles capture the Chamaeleon~I dark cloud. Concerning the Orion cloud complex, data regarding Orion~A have already been collected by \citet{Meingast16}, and Orion~B measurements have been performed as part of the VISTA science verification phase. In addition to the observations of the five star-forming complexes in VISIONS, the deep subsurvey also targets Barnard~59, an actively star-forming cloud at the Galactic northern end of the Pipe~nebula.

Regarding the deep data, the observation setup aims to provide observations of similar or better quality than the Orion~A data from \citet{Meingast16}, especially with respect to the sensitivity and resolution. We map the targeted areas in $J$, $H$, and $K_S$, where all individual tile pointings feature a total on-sky integration time of \SI{600}{\second} in each passband. For scheduling purposes, all deep fields except Barnard~59 in the Pipe~nebula, are imaged twice, each with a \SI{300}{\second} tile exposure time and sky offsets. The purpose of repeated measurements is to facilitate a more reliable construction of a background model, which is a complex process due to the presence of extended emission. The Pipe field is imaged without a sky offset, reaching the target exposure time of \SI{600}{\second} with a single observation block. For all fields, the detector integration time (DIT) is set to \SI{5}{\second} in the $J$ band and \SI{2}{\second} in $H$ and $K_S$ bands. Furthermore, the number of exposures with exposure time DIT (NDIT) is set to 10 in $J$ and 25 in $H$ and $K_S$. With three jitter positions and the obligatory six offsets per tile, the total integration time then amounts to \SI{300}{\second}. The number of individual tiles that include (exclude) Orion amounts to 94 (57) constructed from 2094 (1080) pawprints. Together, these data capture an area of 35.8 $(17.5)\,\si{\deg\squared}$ with a total on-sky exposure time of 9 $(5)\,\si{\hour}$.

\subsubsection{Control subsurvey}
\label{sssec:control_subsurvey}

The purpose of the control subsurvey is to provide data for regions without an excess of extinction to complement the deep fields. Such observations enable the sampling of stellar populations as well as of the extragalactic component to facilitate a statistical comparison between field and star-forming regions. For this reason, the control fields necessarily need to be located at the same Galactic latitude as their science counterparts and should ideally also be relatively close in terms of their Galactic longitude. However, choosing a control field with such characteristics can be especially challenging for regions in the vicinity of the Galactic Plane or projected against the Milky Way bulge. 

Similarly to the deep observations, we inspected the dust emission as captured in the \textit{Planck} \SI{857}{\giga\hertz} band for the control field selections. For Corona Australis and Chamaeleon, their relatively isolated location with respect to other star-forming regions renders the choice trivial, and for Orion, suitable data are already provided by \citet{Meingast16}. For the Ophiuchus and Lupus North deep fields, we choose a region between the two star-forming complexes at about $l\approx345\,\si{deg}$, $b\approx16.5\,\si{deg}$. For these regions, we further include an additional control field since this area still exhibits a moderate amount of extinction and, moreover, is projected against the Upper Scorpius and Upper Centaurus Lupus OB associations. This additional control field, named Ophiuchus South, is taken on a mirrored position in Galactic latitude with respect to the first control field of the region, at $l=348.8\,\si{deg}$, $b=-16.7\,\si{deg}$. By coincidence, this position also constitutes a suitable control field for Corona Australis. The Lupus South control field is pointed toward a relatively extinction-free region in the Galactic West of the Lupus~III cloud. For Barnard~59 in the Pipe~nebula, there is no appropriate region with low extinction. To nevertheless provide a reasonable estimate of the Galactic population, two separate tiles are imaged: a field offset to the Galactic West and another imaging Baade's window close to the Galactic Center.

The observation strategy for the control subsurvey closely follows that of the deep fields. Specifically, we used the same values for DIT, NDIT, and NJITTER, while omitting a sky offset. With this setup, we obtain a total of 942 (864) pawprints that include (exclude) Orion. The 27 (24) tiles capture an area of 16.2~$(14.4)\,\si{deg^2}$ with a total exposure time of 4.33 $(4)\,\si{\hour}$.

\subsection{Scheduling and requirements}
\label{ssec:scheduling_requirements}

For optimal results regarding the wide survey's goal of measuring proper motions, we scheduled the observations so that the individual epochs would be observed with an as long as possible time interval between them. In total, six epochs were planned for each region, and their spacing was adjusted and optimized for the visibility of a region over the course of a year. For Ophiuchus, Lupus, and Chamaeleon, observations were performed during each ESO observing semester. For Corona Australis and Orion, it was necessary to include two epochs in one ESO period, the first at the start and the second toward the end of the corresponding semester.

\begin{figure}
        \centering
    \resizebox{\hsize}{!}{\includegraphics[]{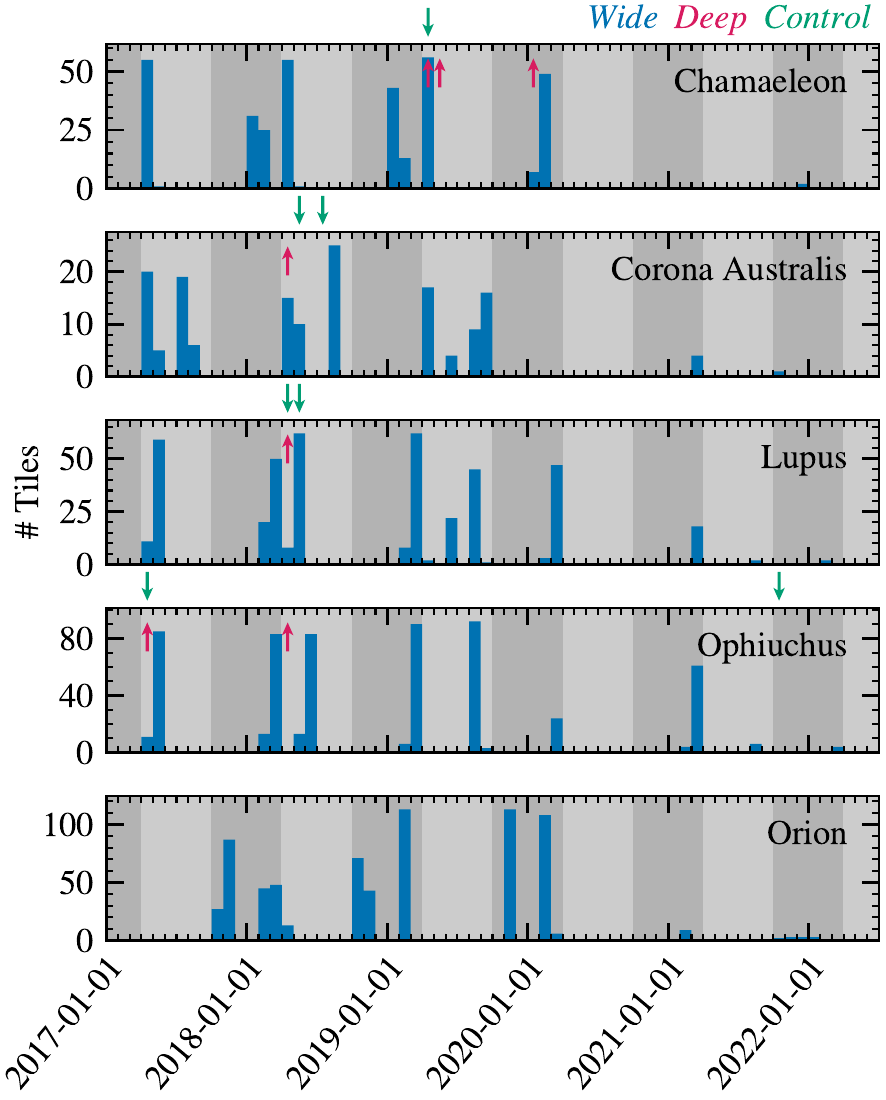}}
        \caption{VISIONS observations over the course of the entire survey. The individual panels show the data collection progress for each separate target region. The blue histograms depict the number of imaged tiles in the wide survey. The observing times for the deep and control surveys are marked with red and green arrows at the top axis of each panel. The alternating gray background refers to individual ESO periods, with period 99 indicating the start of VISIONS observations on April 1, 2017.}
        \label{img:mjd_hist}
\end{figure}

Figure~\ref{img:mjd_hist} shows the distribution of observation times for all tiles throughout the duration of the survey. The blue histogram represents the data collection for the wide survey and observations for the other subsurveys are marked with arrows. The figure highlights consecutive ESO Periods with alternating shades of gray in the background of each panel. Originally, the VISIONS survey was planned to last three years. However, due to incomplete observations and necessary repetitions due to poor data quality, some data were taken after the original survey period ended. Furthermore, during ESO period 105 (1 April to 30 September 2020), as well as for large parts of period 106, no data collection was carried out due to the COVID-19 pandemic. As a consequence, the end-to-end survey duration for VISIONS amounts to almost five years, with the first and last exposures taken on April 1, 2017 and March 27, 2022, respectively.

In contrast to the wide observations, we imposed less strict scheduling requirements on the deep and control subsurveys. For these, we distributed the observations throughout the planned duration of the survey and, when possible, we also included the corresponding control field for each target region in the same observation semester. For Chamaeleon, most of the data were taken in April 2019, with a single observation block, namely, the second execution of the northern field in $J$, imaged in January 2020. For Corona Australis and Lupus, all deep data could be collected over a period of approximately two weeks in April 2018. For Ophiuchus, most of the deep images were taken at the beginning of the survey in April 2017, with two observation blocks in $J$ executed one year later. As already mentioned, all data for Orion within the deep and control subsurveys were part of an earlier publication.

Since the wide survey comprises by far the largest subsurvey and the execution of individual observation blocks was time-critical, we imposed relatively loose constraints on observing conditions. Specifically, we required an airmass of less than 1.7, an image quality of \SI{1}{\arcsec} or better, a minimum angular distance of \SI{30}{deg} to the moon without any restriction on the moon phase, and a thin cirrus (sky transparency variations above \SI{10}{\percent}) was acceptable as well. For the deep and control fields, the same constraints applied to the moon and the airmass. However, image quality was required to be \SI{0.8}{\arcsec} or better, with clear (less than \SI{10}{\percent} of the sky above \SI{30}{deg} elevation covered in clouds; transparency variations under \SI{10}{\percent}) or photometric (no visible clouds; transparency variations under \SI{2}{\percent}) sky observing conditions. Observations were also allowed during twilight for all subsurveys.

VISIONS did not require special calibration data, but it does rely on the standard VISTA calibration plan, as detailed in the VISTA user manual\footnote{\href{https://www.eso.org/sci/facilities/paranal/instruments/vircam/doc.html}{www.eso.org/sci/facilities/paranal/instruments/vircam/doc.html}}. These data include dark frames, twilight flats, and dome flats for determining the non-linearity of the detector array, as well as a dedicated sequence for calculating the read noise and gain levels of the detectors. Furthermore, the survey does not require additional observations to determine zero points because the processing pipeline calibrates all flux measurements with the aid of 2MASS photometry.

\subsection{Data processing, products, and dissemination}
\label{ssec:processing_products_dissemination}

The standard processing pipeline for VIRCAM data, operated by the Cambridge Astronomy Survey Unit (CASU\footnote{\href{http://casu.ast.cam.ac.uk/}{casu.ast.cam.ac.uk}}), is not the optimal choice for surveys aiming to deliver the best possible image quality. As shown by \citet{Meingast16}, the CASU pipeline exhibits two major drawbacks. First, to construct a fully contiguous tile, the images captured by VIRCAM need to be resampled onto a common reference frame. In the CASU pipeline, this process is based on bilinear interpolation, which oftentimes degrades image quality, thus lowering the output resolution by about 20\%. Second, the photometric calibration applies an extinction correction, rendering VIRCAM data regarding star-forming regions provided by CASU difficult to interpret. At the same time, CASU calibrates the data products toward the VISTA-specific photometric system, requiring a color-correction to facilitate a combination with 2MASS photometry. VISIONS strives to deliver both the highest-quality image products and unbiased photometric properties that enable the determination of intrinsic stellar colors. At the same time, our intention is to combine the VISIONS source catalogs with 2MASS, making a strict calibration toward the 2MASS photometric system an indispensable requirement. To achieve these goals, the VISIONS team deploys a dedicated pipeline infrastructure at the University of Vienna, where the core functionality has already been tested by \citet{Meingast16}. A description of the data flow in the pipeline and a detailed view of the functionality of individual modules is available in the second entry in the series of VISIONS articles \citep{VISIONSII}.

Briefly summarized, the VISIONS pipeline produces science-ready images comprising co-added tiles and stacks, as well as corresponding source catalogs. Fluxes are calibrated with respect to 2MASS photometry in the Vega system and \textit{Gaia} Data Release 3 (DR3) is used as an astrometric reference \citep[][]{GaiaDR3,Lindegren21}. For bright sources ($\lesssim 14\,\si{mag}$), photometric uncertainties amount to about \SI{5}{mmag}, while absolute position errors are of the order of \SI{10}{mas} for individual sources on co-added data products. Apart from on-sky positions and source magnitudes, the published data will further include errors on positions and magnitudes, point spread function shape descriptions (e.g., full width at half maxima and ellipticities), and a series of quality flags indicating the reliability of individual measurements. The final product content may vary between different data releases, and detailed descriptions will be published simultaneously with the data products themselves.

The intended release schedule comprises consecutive publications of the data products, each stage introducing all observations for a particular star-forming region. At the time of writing, the VISIONS team has processed and published two data releases, which are available through the ESO Phase 3 Release Manager\footnote{\href{http://eso.org/rm/publicAccess\#/dataReleases}{eso.org/rm/publicAccess\#/dataReleases}}. Data release 1 consists of a small set of products, only including deep data in the Ophiuchus region. Data release 2 contains all observed pointings associated with the Corona Australis molecular cloud complex. For the remaining star-forming regions and the wide field observations in Ophiuchus, we plan to process the observations and publish separate data releases. These future public releases will be made available through the ESO archive, in accordance with the ESO Science Data Products Standard\footnote{\href{https://www.eso.org/sci/observing/phase3/p3sdpstd\_v6.pdf}{www.eso.org/sci/observing/phase3/p3sdpstd\_v6.pdf}}. We note here that these releases do not include proper motion measurements since, for optimal results, these require to also process VHS data in the VISIONS pipeline framework.

\section{Scientific objectives}
\label{sec:scientific_objectives}

The VISIONS program has collected multi-epoch NIR imaging data of five nearby star-forming regions. These data products will facilitate the construction of a sub-arcsecond image atlas in the $J$, $H$, and $K_S$ passbands, accompanied by source catalogs. Additionally, the survey has visited each region at least six times over its runtime, allowing for the computation of precise stellar proper motions, as well as determining long-term variability. We expect that the survey has collected data on about $10^{8}$ individual sources in all star-forming complexes\footnote{The expected number of observed individual sources is an extrapolation from VISIONS observations in the Corona Australis star-forming region.}. For the closest star-forming regions, VISIONS observations are capable of capturing objects with masses only several times that of Jupiter and deeply embedded young stellar objects at a physical resolution of up to \SI{100}{AU}. All data will be made publicly available, allowing the community to address a range of scientific questions.

\subsection{Proper motions of young stars and low-mass objects}
\label{ssec:proper_motions}

\begin{figure}
        \centering
    \resizebox{\hsize}{!}{\includegraphics[]{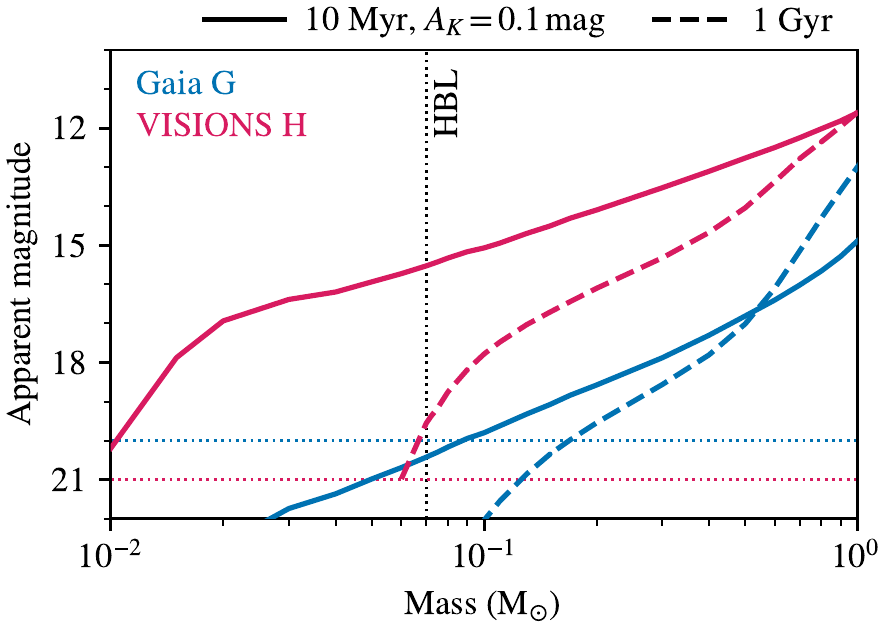}}
        \caption{Apparent magnitudes in the \textit{Gaia} $G$ (blue) and VISIONS $H$ (red) passbands as a function of mass for objects at a distance of \SI{400}{pc}. Solid lines represent a case of young stars with an age of \SI{10}{Myr} located in a region with $A_K=0.1\,\si{mag}$ foreground extinction. The dashed lines represent field stars at an age of \SI{1}{Gyr} without any foreground extinction. The dotted vertical and horizontal lines mark the hydrogen burning limit (HBL) and the surveys' sensitivity limits, respectively.}
        \label{img:bd}
\end{figure}

We have designed the VISIONS observations largely out of a motivation to facilitate proper motion measurements of both the entire population of embedded objects in molecular clouds and optically revealed young and low-mass stars. With this strategy, our survey is to a great extent complementary to the data provided by \textit{Gaia}. Given it carries out observations in the optical wavelength range, \textit{Gaia} is unable to capture the vast majority of embedded sources, either due to large amounts of foreground extinction or because of extinction caused by circumstellar disks or envelopes. Moreover, low-mass objects, such as brown dwarfs, also become increasingly difficult to detect with \textit{Gaia} during their evolution. To visualize this issue, Fig.~\ref{img:bd} shows the apparent magnitude of objects with masses between 0.01 and \SI{1}{M_{\odot}} for two different scenarios for the VISIONS $H$ and \textit{Gaia} $G$ bands in red and blue, respectively. All masses and absolute magnitudes of the objects are taken from \citet{Baraffe15}. In this figure, the solid lines represent a young population (e.g., OB association) at an age of \SI{10}{Myr} and an extinction of $A_K=0.1\,\si{mag}$. The dashed lines denote objects in the Galactic field, without extinction and with an age of \SI{1}{Gyr}. To compute the apparent magnitudes for both cases, we assumed a distance of \SI{400}{pc}, namely, the distance of the farthest star-forming complex in the VISIONS survey, and we used the extinction law of \citet{Wang19}. We also show the hydrogen burning limit (marked with HBL at \SI{0.07}{M_{\odot}}) as a dotted vertical line and the approximate sensitivity limits for both surveys as dotted horizontal lines ($G_{\mathrm{lim}} \approx 20\,\si{mag}; H_{\mathrm{lim}} \approx 21\,\si{mag}$). We note here that the magnitude limit in the \textit{Gaia} survey depends on the scanning law and local source density and can reach up to \SI{21}{mag} for selected areas \citep[e.g.,][]{Cantat-Gaudin22}.
    
Following Fig.~\ref{img:bd}, for the young embedded case (solid lines), \textit{Gaia} reaches its sensitivity limit at about a mass of \SI{0.1}{M_{\odot}}, whereas the VISIONS $H$ observations theoretically are able to recover sources with masses even below \SI{0.01}{M_{\odot}} ($\approx 10\,\si{M_{\mathrm{Jup}}}$). Sources at the hydrogen burning limit of \SI{0.07}{M_{\odot}} are easily detected at magnitudes of approximately $H=16\,\si{mag}$ at this distance. Nevertheless, low-mass objects in particular grow increasingly faint during their evolution. In the second scenario depicted in Fig. \ref{img:bd}, with regard to field stars (dashed lines), sources near the hydrogen burning limit with ages around \SI{1}{Gyr} have an $H$-band magnitude of about \SI{19}{mag} and, at the same time, they exhibit an optical brightness that is fainter by several orders of magnitude -- far below \textit{Gaia}'s sensitivity limit. In this case, \textit{Gaia} is able to provide astrometry for objects with masses of \SI{0.2}{M_{\odot}} or greater, whereas VISIONS still captures objects below the hydrogen burning limit.
    
Based on preliminary tests, including data in the framework of VHS, we were able to to calculate proper motions with a precision of the order of \SI{1}{\mas\per\year} for bright sources ($H \lesssim 16\,\si{mag}$), or the equivalent transverse velocity accuracy of \SI{0.5}{\kilo\meter\per\second} at a distance of \SI{100}{pc}. These numbers are comparable to the precision of the radial velocities of embedded objects from the APOGEE survey \citep[][]{Majewski17} in the Orion star formation region \citep[e.g.,][]{Hacar16,Josefa21}. In addition, VISIONS will enable the combination of its astrometric information with high-precision positional information from radio observations. For embedded sources accessible with VLA, VLBA, or ALMA, absolute astrometry down to the micro-arcsec-level will become available, thereby further boosting the absolute calibration of the astrometric reference frame and improving the already determined proper motions.

\subsection{YSO identification and characterization}
\label{ssec:yso_identification}

In our previous work, we  showed that a redesigned data processing workflow can improve the resolution of VIRCAM data products by approximately 20\%, compared to the products of other pipelines \citep[][]{Meingast16}. As a result of the excellent resolution of the images of that study, it was possible to inspect the morphology of YSO systems in Orion. In particular, the appearance of light scattered off nebulae surrounding young stars proved to be a powerful tool for confirming their protostellar nature. \citet{Josefa19} complemented the VISTA/VIRCAM NIR images with data products from \textit{Spitzer}, \textit{Herschel}, and \textit{WISE} to perform a rigorous search for YSOs in the Orion~A molecular cloud. Their work produced the most complete and reliable sample of YSOs in this region to date and validates the importance of NIR observations for identifying young stars and detecting false positives. Furthermore, high-resolution imaging products will allow for the estimation of various properties of YSOs, such as the inclination of the envelope cavity, which is critical for modeling YSO systems \citep[][]{Robitaille06}.

Outflows and high-velocity bipolar jets are further examples of features made accessible by high-quality NIR imaging data. They are characteristic of the early evolutionary stages of YSOs and represent an important instance of feedback during the process of star formation. The high velocities of these ejections (hundreds of \si{\kilo\meter\per\second}) allow for the study of their morphological evolution and thus the possibility to gather information on the mechanisms responsible for the creation of these objects. In particular, VISIONS and VHS observations provide time baselines of up to \SI{10}{\year} or more that can be used to study outflow progression \citep[e.g.,][]{Reipurth01}. Said baselines could even be extended by another decade by including 2MASS data, although with a significantly reduced image quality.

\subsection{Cluster formation and the initial mass function}
\label{ssec:cluster_formation_IMF}

VISIONS observations are particularly effective in detecting young low-mass objects. VISIONS can observe objects with masses of only a few \si{\mjup} in all cloud complexes included in the survey and, depending on their distance, even at ages above \SI{10}{Myr}. Together with the available astrometric information, the survey enables a robust identification of star formation products across the entire mass spectrum (Fig.~\ref{img:bd}). Furthermore, the detection of free-floating planets is within reach \citep[][]{Caballero18}. As such, with VISIONS it will become possible to more precisely investigate regional variations of the mass function at the lowest substellar masses between and within the star formation sites. At the same time, questions concerning the effects of local feedback on the production of low-mass stars and the connection between the maximum stellar mass and the cluster size can be evaluated.

It is thought that the evolution of star clusters is primarily determined by the early stages of their evolution. In this context, a number of parameters and processes influence the fate of proto-clusters. This includes the number of ejected objects and their masses, mass segregation processes, viralization, the fraction of observed binaries, disk lifetimes, gas expulsion timescales, and the effects of feedback. The quantification of these parameters is within reach with the help of VISIONS data. For example, young wide binary systems are particularly difficult to detect due to blending with fore- or background sources. VISIONS observations are set to enable the determination of young binary systems from both their co-moving nature from proper motions, and spatially resolving the individual components. Moreover, the survey monitors the closest embedded clusters to Earth, providing the best possible physical resolution and allowing constraining model predictions and parameters.

\subsection{3D maps and motion of the interstellar medium}
\label{ssec:3D_motion_ISM}

\begin{figure}
        \centering
    \resizebox{\hsize}{!}{\includegraphics[]{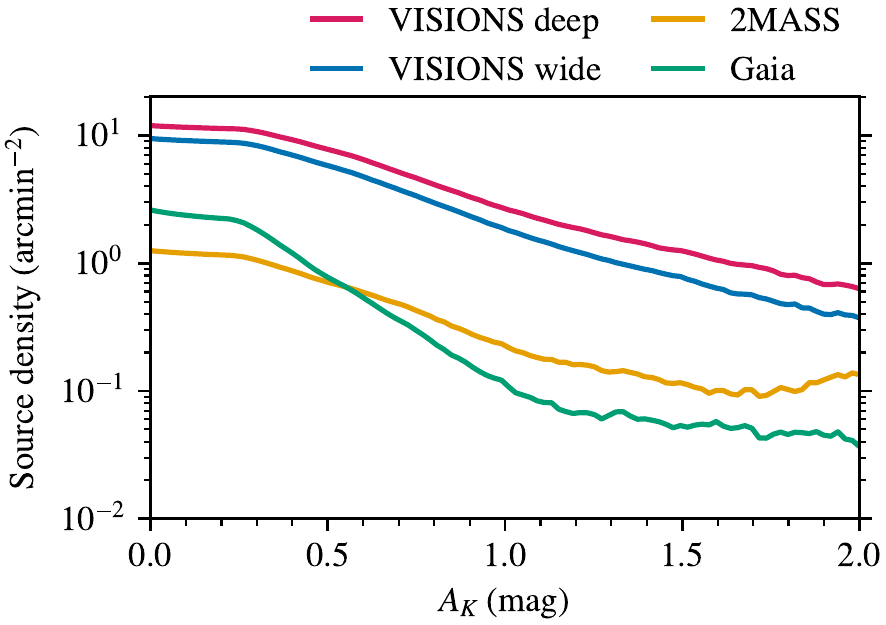}}
        \caption{Source density as function of extinction for the VISIONS, \textit{Gaia}, and 2MASS missions, using data collected on the Orion~A star-forming region. The red and blue lines represent VISIONS wide and deep data, while the green and yellow lines mark the \textit{Gaia} and 2MASS surveys, respectively. Across all levels of extinction, VISIONS outperforms the other surveys by up to an order of magnitude.}
        \label{img:source_density}
\end{figure}

Following the advent of \textit{Gaia}, it has become possible to construct 3D maps of the dust distribution in the nearby interstellar medium \citep[e.g.,][]{Green18, Lallement19, Green19, Leike20, Lallement22}. Prior to the availability of \textit{Gaia} parallaxes on large scales, \citet{Green16} had already constructed a 3D map of the dust distribution using Pan-STARRS \citep[][]{Kaiser10} and 2MASS photometry, thereby demonstrating that stellar photometry plays a vital role in correctly inferring distances to stars. However, even the most current 3D dust maps, constructed with \textit{Gaia}, only manage to accurately map the low-density ISM and the envelopes of star-forming complexes. With the upcoming data releases, VISIONS will provide an accurate census of the embedded and background populations of nearby cloud complexes, thus complementing \textit{Gaia} and enabling a more robust mapping of high-density ISM environments. To visualize the gain in available background sources that can be used for measuring reddening due to dust extinction, Fig.~\ref{img:source_density} displays the source density as a function of the foreground extinction for VISIONS, 2MASS, and \textit{Gaia} data. We calculate the densities in a field that covers about \SI{20}{deg^2} of the Orion~A molecular cloud, as published by \citet{Meingast16}. \textit{Gaia} manages to capture about twice as many sources as 2MASS at low extinctions, and only at $A_K=0.5\,\si{mag}$ does 2MASS provide better completeness than \textit{Gaia}. Nevertheless, VISIONS outperforms both other surveys by almost an order of magnitude across all extinction levels.

Furthermore, recent studies have managed to determine the 3D motion of nearby star-forming gas. Specifically, \citet{Josefa21} used \textit{Gaia} astrometry and CO gas radial velocities to establish a connection between a subset of the embedded YSOs in the Orion star-forming complex and the local 3D gas motion. One major constraint of their study was the limited availability of proper motion measurements of embedded stars from \textit{Gaia}. VISIONS will enable measuring the proper motions of young embedded objects, which, together with infrared radial velocities (from e.g., APOGEE), will permit a far more reliable determination of the 3D gas motion in the nearby ISM.

\subsection{Properties of dense cores}
\label{ssec:dense_cores}

VISIONS deep observations will provide source catalogs built from long integration times, reaching about \SI{6}{mag} fainter sensitivity limits than 2MASS. These observations, combined with data from \textit{WISE} and \textit{Spitzer}, will provide the best census of stars so far in the background of dense cores. As a consequence, the construction of extinction maps, reaching resolutions of 10 -- \SI{15}{\arcsec} towards the bulge and of 30--\SI{60}{\arcsec} towards the Galactic anticenter will become possible \citep[e.g.,][]{Meingast18}. These extinction maps will be free from bias due to dust temperature, $\beta$ variations, gas excitation conditions, or molecular depletion. In combination with \textit{Herschel} dust emission measurements, we expect to be able to construct a well-sampled core mass function that can be compared with its stellar counterpart. Also, VISIONS will enable comparing starless and star-forming cores, establishing a connection between dense cores and YSOs, and mapping their distribution in the ensemble of clouds to correlate them with the distribution of YSOs. 

\subsection{Dust properties and reddening law}
\label{ssec:dust_reddening}

The NIR wavelength regime is ideally suited for inferring the column-density distribution of highly extincted areas. For such applications, extinction mapping has not only been shown to be the most reliable method \citep{Goodman09}, but also the reddening law in the NIR itself appears to show variations of only a few percent in different environments \citep[][]{Meingast18}.

VISIONS will enable further testing of the universality of the NIR reddening law, since the observed dust clouds are mostly isolated and allow for an unbiased investigation free from background confusion. Furthermore, the combination of extinction mapping with submillimeter dust emission data obtained with \textit{Herschel} and \textit{Planck} will also provide constraints on the ratio of the submillimeter dust opacity and the NIR extinction coefficient. Where available, VISIONS, in combination with ground-based bolometer data, would also allow evaluating the dust opacity spectral index $\beta$ \citep[e.g.,][]{Forbrich15}.


\section{Summary}
\label{sec:summary}

This manuscript serves as the main reference to the ESO public survey VISIONS: The VISTA Star Formation Atlas. The survey targets five nearby star-forming regions, namely Ophiuchus, Lupus, Chamaeleon, Corona Australis, and Orion. VISIONS observations have been carried out with the VISTA telescope and its integrated NIR camera VIRCAM, located at the ESO Paranal observatory in Chile. The observation schedule, starting in April~2017 and lasting until March 2022, was split into three different subsurveys: the wide, deep, and control programs. They are targeted at collecting wide-field multi-epoch observations in the $H$ band, highly sensitive $JHK_S$ imaging data of highly extincted areas, and $JHK_S$ data on areas largely free of extinction, respectively. While the wide observations facilitate proper motion measurements of objects beyond the reach of the ESA \textit{Gaia} mission, the deep and control data provide a detailed view of the embedded populations. Furthermore, VISIONS has been designed to be complementary to the first generation ESO VISTA public survey VHS, which provides the $J$ and $K_S$-band observation for the wide subsurvey. Additionally, using VHS extends the available time baseline for proper motion measurements up to \SI{10}{yr} or more.

The main objective of VISIONS is to establish a legacy NIR archive that will allow the community to improve our understanding of the process of star formation. VISIONS will facilitate addressing a range of scientific topics, including the determination of the kinematics of embedded objects down to substellar masses, the reliable identification and characterization of YSOs, the construction of an unbiased view of the formation and evolution of embedded clusters, the verification of the products of star formation in relation to the initial mass function, the support of the creation of 3D dust maps and the deduction of 3D dynamics of the local interstellar medium, as well as the reliable investigation of dust properties and the reddening law. To achieve these goals, VISIONS is set to join ranks with prominent surveys and missions, such as 2MASS, APOGEE, \textit{Spitzer}, \textit{WISE}, ALMA, and \textit{Gaia}, to deliver the most accurate view of the star formation processes in the solar neighborhood.

\begin{acknowledgements}
We thank the anonymous referee for the useful comments that helped to improve this publication.
We thank the ESO Survey Team and the Archive Science Group for their helpful and constructive feedback and collaboration during the survey's preparation, execution, and data publication phases.
This research has used the services of the ESO Science Archive Facility.
This research has made use of Astropy\footnote{\href{http://www.astropy.org}{www.astropy.org}}, a community-developed core Python package for Astronomy \citep{astropyI, astropyII}.
This research has made use of "Aladin sky atlas" developed at CDS, Strasbourg Observatory, France \citep{aladin} and the table processing tools TOPCAT and STIL \citep{topcat}.
We also acknowledge the various Python packages that were used for the preparation of this manuscript, including NumPy \citep{numpy}, SciPy \citep{scipy}, scikit-learn \citep{scikit-learn}, scikit-image \citep{scikit-image}, and Matplotlib \citep{matplotlib}.
This research has made use of the SIMBAD database operated at CDS, Strasbourg, France \citep{simbad}. This research has used the VizieR catalog access tool, CDS, Strasbourg, France (\citealp{vizier}; DOI: \href{dx.doi.org/10.26093/cds/vizier}{10.26093/cds/vizier}).
This work has made use of data from the European Space Agency (ESA) mission
{\it Gaia} (\href{https://www.cosmos.esa.int/gaia}{www.cosmos.esa.int/gaia}), processed by the {\it Gaia}
Data Processing and Analysis Consortium (DPAC,
\href{https://www.cosmos.esa.int/web/gaia/dpac/consortium}{www.cosmos.esa.int/web/gaia/dpac/consortium}). Funding for the DPAC
has been provided by national institutions, in particular the institutions
participating in the {\it Gaia} Multilateral Agreement.
A.B. acknowledges partial funding by the Deutsche Forschungsgemeinschaft Excellence Strategy - EXC 2094 - 390783311 and the ANID BASAL project FB210003.
A.H. acknowledges the support and funding from the European Research Council (ERC) under the European Union’s Horizon 2020 research and innovation programme (Grant agreement Nos. 851435).
D.M. acknowledges support from ANID Basal project FB210003.
K.M. acknowledges funding by the Science and Technology Foundation of Portugal (FCT), grants No. PTDC/FIS-AST/7002/2020 and  UIDB/00099/2020.
K.P.R. acknowledges support from ANID FONDECYT Iniciaci\'on 11201161.
\end{acknowledgements}

\clearpage
\bibliography{references.bib}

\end{document}

%% file: table_setup.tex
\begin{sidewaystable*}
    \caption{VISIONS observation setup. This table contains data on the number of images, data volume, on-sky area, and the imaging setup for each sub-survey.}
    \label{tab:overview}
    \begin{tabular*}{\linewidth}{l c c c c c c c c c c c c c c}
        \hline\hline
Region	&	$l$	&	$b$	&	Tiles	&	Tiles	&	Pawprints	&	Volume	&	Area	&	DIT	&	NDIT	&	NJITTER	&	Exp. time	&	Exp. time	&	Sky offset	&	Filter	\\
	&		&		&	single	&	total	&	total	&		&		&		&		&		&	single	&	total	&		&		\\
	&	(deg)	&	(deg)	&	(\#)	&	(\#)	&	(\#)	&	(GB)	&	(deg$^2$)	&	(s)	&	(\#)	&	(\#)	&	(s)	&	(h)	&		&		\\
\\																													
\multicolumn{15}{c}{$Wide$}	\\																												
\hline																													
Chamaeleon	&	298.6	&	-16.1	&	51	&	306	&	9180	&	2464	&	84.5	&	3	&	2	&	5	&	60	&	5.1	&	\xmark	&	$H$	\\
Chamaeleon (Musca)	&	301.5	&	-7.7	&	5	&	30	&	900	&	242	&	8.8	&	3	&	2	&	5	&	60	&	0.5	&	\xmark	&	$H$	\\
Corona Australis	&	360.0	&	-18.8	&	25	&	150	&	4500	&	1208	&	41.4	&	3	&	2	&	5	&	60	&	2.5	&	\xmark	&	$H$	\\
Lupus	&	337.9	&	12.3	&	70	&	420	&	12600	&	3382	&	115.9	&	3	&	2	&	5	&	60	&	7	&	\xmark	&	$H$	\\
Ophiuchus	&	353.3	&	18.4	&	96	&	576	&	17280	&	4639	&	159.0	&	3	&	2	&	5	&	60	&	9.6	&	\xmark	&	$H$	\\
Orion	&	209.9	&	-19.8	&	114	&	684	&	20520	&	5508	&	188.8	&	3	&	2	&	5	&	60	&	11.4	&	\xmark	&	$H$	\\
Epoch $X$	&	-	&	-	&	20	&	120	&	3600	&	966	&	-	&	3	&	2	&	5	&	60	&	2	&	\xmark	&	$H$	\\
\hline																													
Subtotal	&		&		&	381	&	2286	&	68580	&	18409	&	598.4	&		&		&		&		&	36.1					\\
\hline																													
\\																													
\multicolumn{15}{c}{$Deep$}	\\																												
\hline																													
Chamaeleon	&	296.9	&	-15.5	&	2	&	12	&	216	&	58	&	3.5	&	5/2/2	&	10/25/25	&	3	&	300	&	1	&	\cmark	&	$JHK_S$	\\
Corona Australis	&	359.9	&	-17.8	&	1	&	6	&	108	&	29	&	1.8	&	5/2/2	&	10/25/25	&	3	&	300	&	0.5	&	\cmark	&	$JHK_S$	\\
Lupus North	&	339.0	&	16.2	&	1	&	6	&	108	&	29	&	1.8	&	5/2/2	&	10/25/25	&	3	&	300	&	0.5	&	\cmark	&	$JHK_S$	\\
Lupus South	&	339.5	&	9.3	&	1	&	6	&	108	&	29	&	1.8	&	5/2/2	&	10/25/25	&	3	&	300	&	0.5	&	\cmark	&	$JHK_S$	\\
Ophiuchus	&	353.8	&	16.8	&	4	&	24	&	432	&	116	&	6.8	&	5/2/2	&	10/25/25	&	3	&	300	&	2	&	\cmark	&	$JHK_S$	\\
Orion (w/ sky)	&	209.1	&	-19.4	&	2	&	10	&	312	&	84	&	3.5	&	5/2/2	&	9/17/15	&	6/5/5	&	540/340/300	&	1.0	&	\cmark	&	$JHK_S$	\\
Orion (w/o sky)	&	211.5	&	-19.4	&	9	&	27	&	702	&	188	&	14.8	&	5/2/2	&	8/27/20	&	3/5/5	&	240/540/400	&	3.0	&	\xmark	&	$JHK_S$	\\
Pipe	&	357.4	&	6.9	&	1	&	3	&	108	&	29	&	1.8	&	5/2/2	&	10/25/25	&	6	&	600	&	0.5	&	\xmark	&	$JHK_S$	\\
\hline																													
Subtotal	&		&		&	21	&	94	&	2094	&	562	&	35.8	&		&		&		&		&	9.0					\\
\hline																													
\\																													
\multicolumn{15}{c}{$Control$}	\\																												
\hline																													
Chamaeleon	&	308.7	&	-15.6	&	1	&	3	&	108	&	29	&	1.8	&	5/2/2	&	10/25/25	&	6	&	600	&	0.5	&	\xmark	&	$JHK_S$	\\
Corona Australis	&	3.6	&	-17.9	&	1	&	3	&	108	&	29	&	1.8	&	5/2/2	&	10/25/25	&	6	&	600	&	0.5	&	\xmark	&	$JHK_S$	\\
Lupus North	&	344.1	&	16.2	&	1	&	3	&	108	&	29	&	1.8	&	5/2/2	&	10/25/25	&	6	&	600	&	0.5	&	\xmark	&	$JHK_S$	\\
Lupus South	&	333.1	&	9.2	&	1	&	3	&	108	&	29	&	1.8	&	5/2/2	&	10/25/25	&	6	&	600	&	0.5	&	\xmark	&	$JHK_S$	\\
Ophiuchus North	&	345.3	&	16.8	&	1	&	3	&	108	&	29	&	1.8	&	5/2/2	&	10/25/25	&	6	&	600	&	0.5	&	\xmark	&	$JHK_S$	\\
Ophiuchus South	&	348.8	&	-16.7	&	1	&	3	&	108	&	29	&	1.8	&	5/2/2	&	10/25/25	&	6	&	600	&	0.5	&	\xmark	&	$JHK_S$	\\
Orion	&	233.3	&	-19.4	&	1	&	3	&	78	&	21	&	1.8	&	5/2/2	&	8/27/20	&	3/5/5	&	240/540/400	&	0.33	&	\xmark	&	$JHK_S$	\\
Pipe North	&	352.1	&	6.6	&	1	&	3	&	108	&	29	&	1.8	&	5/2/2	&	10/25/25	&	6	&	600	&	0.5	&	\xmark	&	$JHK_S$	\\
Pipe South	&	1.2	&	-3.8	&	1	&	3	&	108	&	29	&	1.8	&	5/2/2	&	10/25/25	&	6	&	600	&	0.5	&	\xmark	&	$JHK_S$	\\
\hline																													
Subtotal	&		&		&	9	&	27	&	942	&	253	&	16.2	&		&		&		&		&	4.33					\\
\hline																													
\textbf{Total}	&		&		&	\textbf{411}	&	\textbf{2407}	&	\textbf{71616}	&	\textbf{19224}	&	\textbf{650.4}	&		&		&		&		&	\textbf{49.4}					\\
\hline																													
\end{tabular*}
\end{sidewaystable*}